# Single-shot multispectral quantitative phase imaging using deep neural network

**Sunil Bhatt[1], Ankit Butola[2,*], Anand Kumar[1], Pramila Thapa[1], Akshay Joshi[3], Neetu Singh[3], Krishna Agarwal[2,#] And Dalip Singh Mehta[1,%]**

[1]Bio-photonics and Green-photonics Laboratory, Department of Physics, Indian Institute of Technology Delhi, Hauz-Khas, New Delhi 110016, India.
[2]Department of Physics and Technology, UiT, The Arctic University of Norway, Norway.
[3] Centre for Biomedical Engineering, Indian Institute of Technology Delhi, Hauz-Khas, New Delhi 110016, India.
[*]ankit.butola@uit.no
[#]krishna.agarwal@uit.no
[%] Dalip.Singh.Mehta@physics.iitd.ac.in

## Abstract

Multi-spectral quantitative phase imaging (MS-QPI) is a cutting-edge label-free technique to determine the morphological changes, refractive index variations and spectroscopic information of the specimens. The bottleneck to implement this technique to extract quantitative information, is the need of more than two measurements for generating MS-QPI images. We propose a single-shot MS-QPI technique using highly spatially sensitive digital holographic microscope assisted with deep neural network (DNN). Our method first acquires the interferometric datasets corresponding to multiple wavelengths ($\lambda$=532, 633 and 808 nm used here). The acquired datasets are used to train generative adversarial network (GAN) to generate multi-spectral quantitative phase maps from a single input interferogram. The network is trained and validated on two different samples, the optical waveguide and a MG63 osteosarcoma cells. Further, validation of the framework is performed by comparing the predicted phase maps with experimentally acquired and processed multi-spectral phase maps. The current MS-QPI+DNN framework can further empower spectroscopic QPI to improve the chemical specificity without complex instrumentation and color-cross talk.

## Introduction

Quantitative phase imaging (QPI) is a prominent label-free technique to measure morphological changes of the biological cells which are transparent in nature[1-3]. Over the past few decades, QPI has been developed both experimentally and computationally for the accurate measurement and classification of various parameters of biological cells/tissues such as surface profile, refractive index, dry mass density, cell membrane fluctuations and among others[4-9]. In QPI, morphological and dynamical information about the samples can be extracted by measuring the path length shift ($\Delta\phi$) associated with the objects which contains combine information of refractive index ($n_s$) and thickness (d) of the specimens. Methods have been developed to improve specificity of QPI system by decoupling the refractive index and thickness of the object[7, 10]. However, these methods have not been widely used due to either experimental complexity or imperfect decoupling the phase map. Nonetheless, improvement of the specificity of QPI requires image acquisition with additional dimensions of measurements, such as sample rotation, angle and wavelength.

In the past, different methods such as diffraction phase microscopy[11, 12], quantitative dispersion microscopy[13] , quantitative phase spectroscopy[14], and dynamic spectroscopic phase microscopy[15] have been developed for phase measurement at multiple wavelengths. These methods have been utilized to overcome issues such as phase noise, phase unwrapping and determination of refractive index dependence on the illumination wavelength. The acquisition of such multi spectral (MS)-QPIs can be done in two ways: 1) sequential illumination mode[16, 17], in which interferograms with different wavelengths are recorded one after the other, and 2) simultaneous illumination mode[18, 19], in which all interferograms are recorded in one go using an RGB color camera. The first method requires manual or mechanical switching of the light source, whereas the second method requires computational separation of the RGB image into different wavelength channels. Computational separation of spectral components suffers from color-cross talk problems due to color cameras[20]. In addition, these methods cannot utilize the information-rich near infrared (NIR) spectrum of biological samples. Although, the broadband light sources such as LEDs, white light sources and super luminescent diodes (SLD's) are also used for MS-QPI, these sources are inappropriate for wavelength dependent applications due to the temporally and spatially low coherence properties and wide spectral bandwidth. In addition, such systems require extra arrangement of dispersion compensation mechanism due to dispersion artefacts, which makes them bulky and costly[21]. Therefore, advance techniques are required to cover wide spectral range, and non-mechanical or manual switching of optical elements for fast and cost-effective study of biological specimens[21].

In the current study, we present a single-shot MS-QPI technique assisted with deep neural network (DNN) to achieve multispectral phase images of industrial and biological samples. For this purpose, we have used a partially spatially coherent (PSC) light based digital holographic microscopy system to obtain multi spectral phase maps from a single input interferogram. The interferometric images are acquired using Linnik interferometer sequentially by wavelength switching and processed for the extraction of phase maps of the sample for each wavelength respectively. These phase maps are used to train the generative adversarial network (GAN). The network is first trained and optimized to predict the phase maps for three different wavelengths from a single input interferogram. We showed examples of the training aspects using a more controlled non-biological sample set comprising of data of optical waveguides whose material and geometry compositions are well-qualified. Further, we showed its utility on biological cells of a particular kind (MG63 osteosarcoma cell). The phase maps retrieved from the experimentally recorded interferograms for both the samples was compared with the train network-generated phase maps to evaluate the performance of the network Therefore, our framework performs both reconstruction as well as multi-spectral estimation in an integrated manner without resorting to the conventional reconstruction approach. This implies that our network learns the physics of reconstruction and encodes the spectral properties of piece-wise homogeneous materials in the overall inhomogeneous samples. Indeed, it assumes that the training dataset has sufficient representatives of the structural and spectral variation expected in the sample. However, it reduces the experimental and computational demands of performing MS-QPI in the conventional manner.

## Experimental Setup:

The experimental setup of the proposed framework to acquire the single-shot MS-QPI is shown in Figure 1. Highly spatially coherent light sources such as lasers cannot be directly use for MS-QPI due to speckles and formation of spurious and parasitic fringes due to reflections from multiple layers of sample or the optical components[16, 22, 23]. These speckles and spurious fringes reduce the resolution as well as the feature of the reconstructed phase maps[16, 24]. Therefore, we realize a partially spatially coherent (PSC) light source i.e., temporally high but spatially low coherent, by passing the laser light through a rotating diffuser (RD) and a multi-multimode fibre bundle (MMFB). The RD scatters photons into various directions and makes temporally narrow and wide angular spectral source, which is further coupled into the input of the MMFB. The effect of spatial, temporal and angular diversity has been utilized to significantly reduce the speckles of the coherent light sources[24].

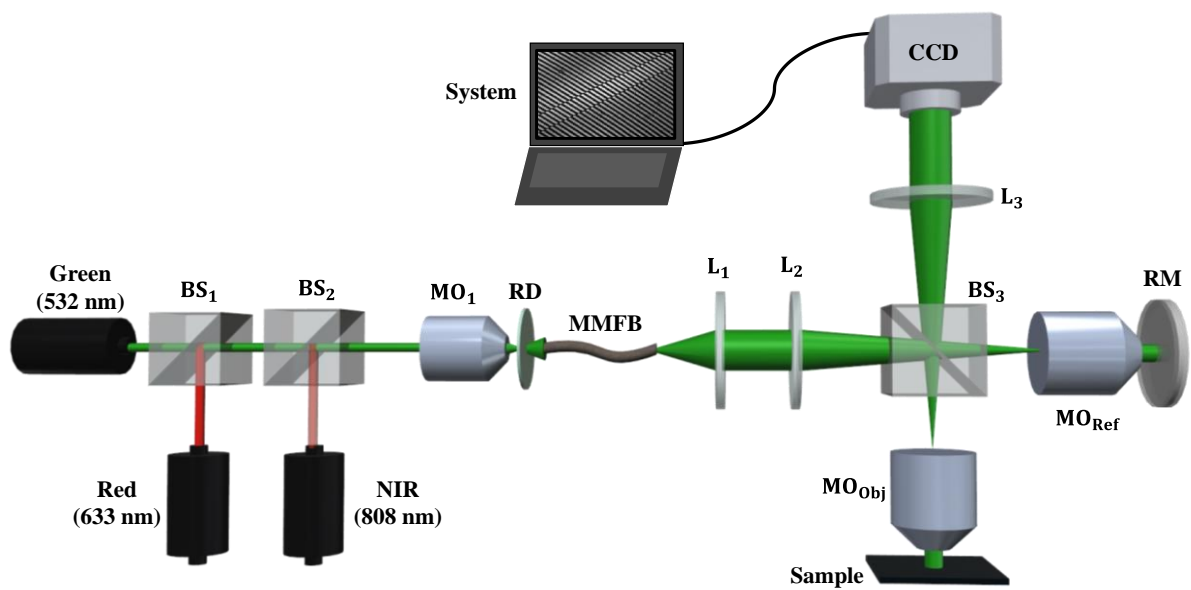

**FIGURE 1** Schematic diagram of multispectral quantitative phase imaging (MS-QPI) system. Three different LASERs (green DPSS ($\lambda$=532 nm), He-Ne ($\lambda$=633 nm), and near infrared ($\lambda$=808 nm)). $BS_1$, $BS_2$ and $BS_3$: beam splitter; MO: microscopic objectives (10X, 0.25 NA); RD: rotating diffuser; MMFB: multi multimode fibre bundle; $L_1$, $L_2$ and $L_3$: lenses; RM: reference mirror; CCD: charge-coupled device (INFINITY2, Lumenera).

Here, we use laser light sources of three different wavelengths, namely green (DPSS, $\lambda_1$ = 532.8 nm.), red (He-Ne laser, $\lambda_2$ = 632.8 nm.), and NIR (laser diode dot module, $\lambda_3$ = 808 nm.). For speckle free imaging, these sources are sequentially passed through a microscopic objective ($MO_1$), RD and MMFB. The light from the MMFB is coupled at the input port of the Linnik interferometer. The output beam from MMFB is collected by the lens $L_1$ which is first collimated and then focused by lens $L_2$ into the back focal plane of two identical microscope objectives $MO_{obj}$ (sample arm) and $MO_{ref}$ (reference arm) by using the 50/50 beam splitter ($BS_3$). The focused light is passed through the $MO_{obj}$ which uniformly illuminates the sample. The back reflected light beam from the sample that holds the sample information interferes with the reference beam at the beam splitter plane. The interference signal is further collimated and projected into the charged coupled device (CCD) camera plane using the tube lens $L_3$. The light sources are switched to record multi-wavelength interferograms, which are used for phase reconstruction of the sample using the Fourier transform (FT) algorithm. The high fringe density interferograms at the camera plane are achieved by tilting the reference mirror (RM). The approach of this study is to generate simultaneous multi spectral phase maps from a single input interferogram. To achieve this purpose, we train a generative adversarial network (GAN) which is basically a DNN for the execution of the study.

**The Architecture of Generative Adversarial Network (GAN):**

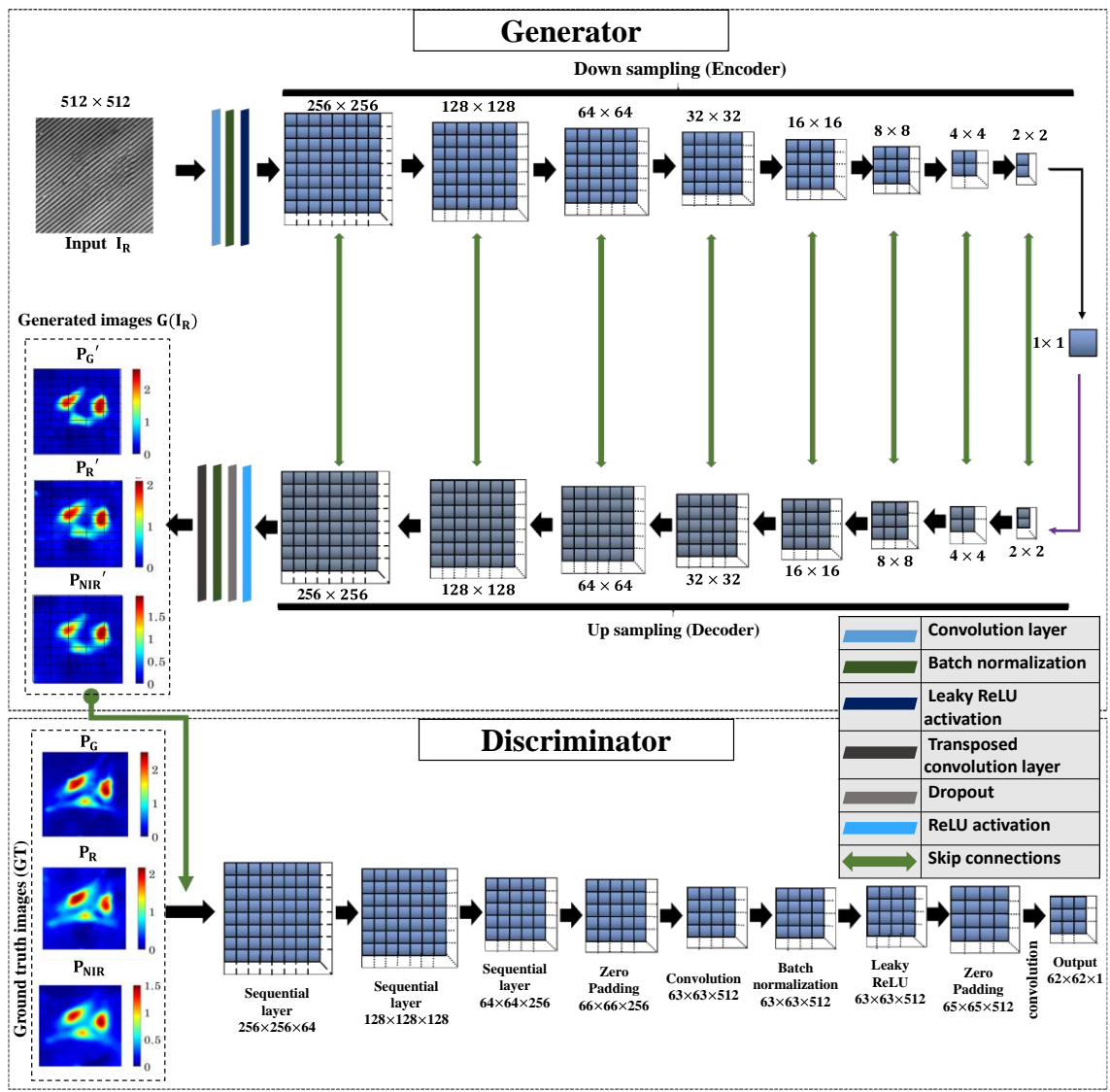

**FIGURE 2** The architecture of generative adversarial network (GAN) to generate multispectral quantitative phase images.

GAN is made up of two models: Generator (G) and Discriminator (D)[25]. The generator creates some mapped images $G(I_R)$, and the discriminator compares them to the ground truth (target) images. Ground truth (GT) images are final output images that have been experimentally processed and then sent into the network for training. The training of these two models and training datasets determines GAN's overall performance. An increase in the amount of training datasets leads to improved network training. GAN's architecture is depicted in Fig. 2. In Fig. 2 we showed the U-net architecture of generator[25, 26] which comprises encoders and decoders and PatchGAN architecture of discriminator[25, 27]. The detailed information about the working of GAN can be found in the supplementary file.

### Cell culturing process of MG63 osteosarcoma cells:

The preparation of MG63 osteosarcoma cells was carried out at Centre for Biomedical Engineering, IIT Delhi, India and optical waveguide was developed in UiT, The Arctic University of Norway. The MG63 osteosarcoma cells are malignant bone cancer cells with abnormality in its cellular structure. The cells were cultured using high-glucose DMEM media with 10% fetal bovine serum (FBS) and 1% antibiotic antimycotic solution and incubated in a humidified atmosphere containing of 95% air and 5% $CO_2$ at 37°C. Before seeding, silicon wafers were sterilized using UV and ethanol treatment for 10 min each. Cells were seeded on sterilized silicon wafer in a 24 well plate and incubated for more than 48 hours in a $CO_2$ incubator. Later, the samples were washed gently with phosphate-buffered saline (PBS) and fixed with 4% paraformaldehyde. To perform the experiment, PBS dipped cells were placed under the proposed MS-QPI system for data recording.

### Workflow of the Framework: Data acquisition, network training, and network testing

The workflow of the present MS-QPI+DNN framework for generation of simultaneous multispectral phase maps from a single input interferogram (i.e., single-shot MS-QPI) is shown in Fig. 3. The current framework is composed of three different parts: data acquisition & processing, training, and testing of the network.

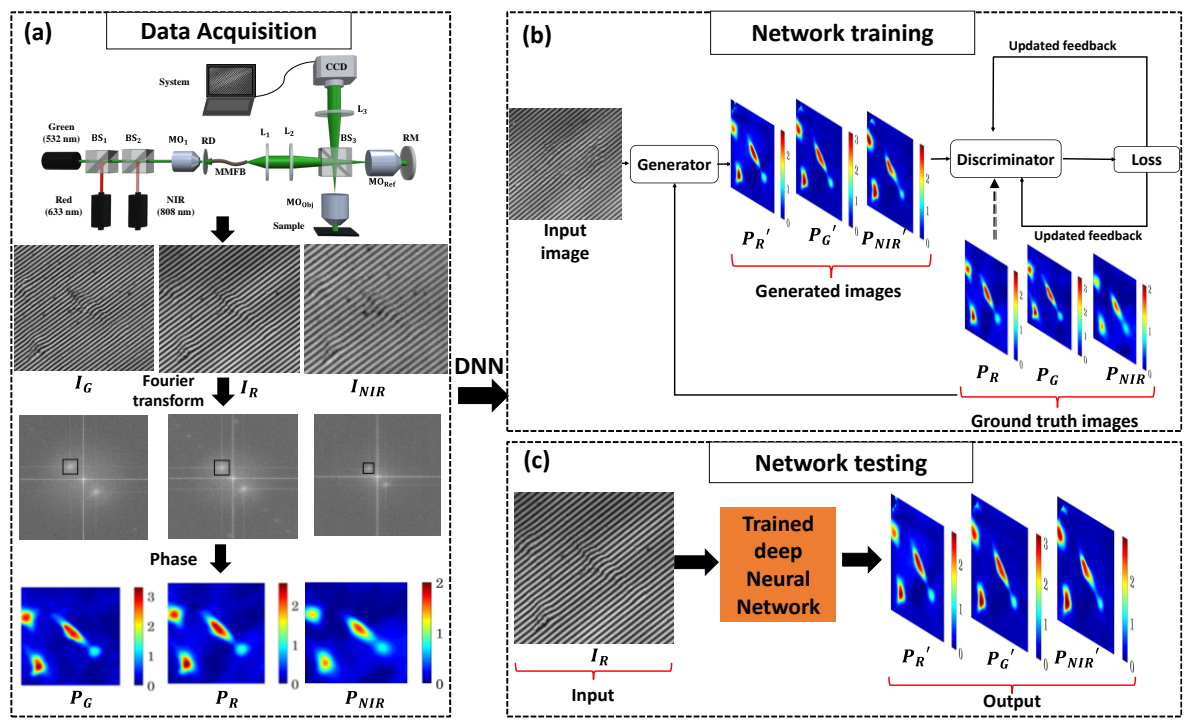

**FIGURE 3** Workflow of the MS-QPI+DNN framework: (a) data acquisition process where $I_G$, $I_R$, $I_{NIR}$, and $P_G$, $P_R$, $P_{NIR}$, are the three wavelength interferograms and their corresponding phase maps for λ=532.8nm. (green), λ = 632.8nm. (red), and λ=808nm. (NIR), respectively using MS-QPI system. (b) block diagram for the training process of DNN. (c) block diagram of network testing for multispectral phase map generation from a single input interferogram.

During data acquisition, we have recorded multi spectral interferograms by switching the light sources as shown in Fig. 3(a). The optical waveguide and MG63 cell data are acquired using the MS-QPI system by employing two identical microscope objectives ($MO_{obj}$=$MO_{ref}$). The recorded interferometric images of samples are reconstructed using Fourier transform (FT) and TIE unwrapping algorithm. For the training of the network, we have used a single input interferogram and three phase maps (ground truth images) corresponding to different $λ_i$ as shown in Fig. 3(b). During training, single interferogram is taken as an input of the generator to generate some mapped images. The discriminator discriminates between the generated and the ground truth images and calculate a loss which can be feed back to the generator and discriminator for fine tuning of the network. After a number of iterations, the generator has enough training to generate same images as that of ground truth i.e., loss will be minimum. The trained network is used for testing as shown in Fig. 3(c).

Initially for the validation of our approach, we use only two wavelength datasets for the optical waveguide i.e., green and red. After successful adaptation of the network with two wavelengths for the optical waveguide, we use three wavelengths i.e., green, red and NIR for the biological sample. The model is trained with 4080 data items for input image size of 256×256 and buffer size of 400 (for data randomization) with 60 epochs for the optical waveguide datasets. For complex structures like MG63 cell datasets, 11,556 data items are employed to train the network. This model is trained for input image size of 512×512 and buffer size of 400 with 70 epochs. The network is programmed in Python (version 3.8.0) and implemented using TensorFlow-GPU (version 2.6) and Keras (version 2.2.0) library functions on Anaconda Jupyter notebook.

### Result and Discussion:

We have shown the comparison between the experimentally processed phase maps and the network generated results of the optical waveguide and MG63 cell dataset. The network is optimized and trained for both the datasets individually. Once, the network is trained, the new test images are used and compared with the experimentally acquired and processed datasets.

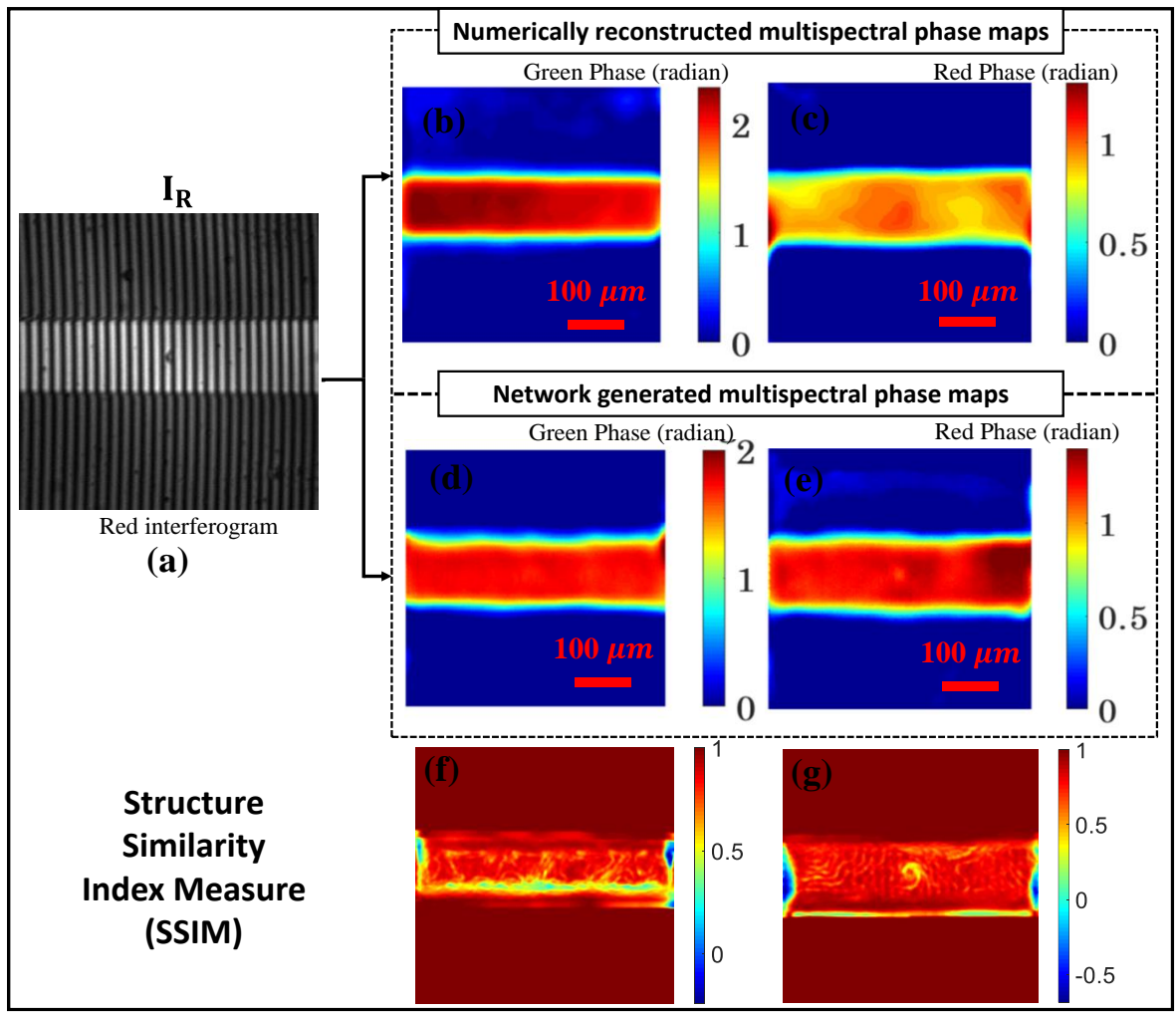

**FIGURE 4** Illustrating comparison between numerically (FT+ TIE algorithm based) reconstructed and network generated multispectral phase map data from a single input interferogram on optical waveguide. (a) is the experimentally recorded input red wavelength interferogram (b), (c) are the numerically reconstructed phase maps for experimentally recorded green and red wavelength interferograms, respectively. (d), (e) are the network generated phase maps corresponding to green and red

wavelength, respectively. (f), (g) represents structured similarity index measure (SSIM) between numerically reconstructed and trained network generated multispectral phase maps. (f) showing the SSIM for green wavelength phase maps between (b) and (d) which is found to be 0.93. (g) showing the SSIM for red wavelength phase maps between (c) and (e) which is found 0.91.

Figure 4 shows the comparison of numerically (FT+ TIE algorithm based) reconstructed and network generated multispectral phase map data of the optical waveguide. The input interferogram corresponding to λ=632.8 nm is shown in Fig. 4 (a). Figure 4 (b) and (c) are numerically reconstructed phase maps corresponding to λ=532.8 nm and λ=632.8nm, respectively for experimentally recorded interferograms. Figure 4 (d) and (e) are the network generated phase maps corresponding to λ=532.8nm and λ=632.8nm, respectively from a single input interferogram $I_R$. The comparison shows that there is a wavelength dependent phase map variation predicted by the network similar as the experimental reconstructed phase maps. Since the phase map is inversely proportional to the illumination wavelength, therefore the phase value decreases as wavelength increases. This is also predicted by the network which can be seen in the color bar, signifies training of the wavelength dependency by the network for phase prediction. In order to quantify the performance of MS-QPM+DNN, structure similarity index measure (SSIM) is calculated to measure the similarity between numerically reconstructed and network generated phase maps[28]. The SSIM values lies in [-1 1][28], where 1 represents the ideal case of identical numerically reconstructed and network generated phase map results while -1 represents dissimilarity in structure between numerically reconstructed and network generated phase map results. Figure 4 (f) shows the SSIM map of phase maps corresponds to λ=532.8nm between Fig. 4 (b) and (d). The net SSIM value is 0.93 which signifies that similarity in structure between the network generated and numerically reconstructed phase maps are in good resemblance. Figure 4 (g) shows the SSIM of phase maps corresponds to λ=632.8nm between Fig. 4 (c) and (e), and the net SSIM is 0.91. This signifies that the similarity in structure between the network generated and numerically reconstructed phase maps are in good resemblance. Hence, the network is well trained for the MS-QPI with higher similarity in structure for multi spectral data generation. The slight mismatch in SSIM value between numerically reconstructed and network generated results may occur due to the network artefacts during data training such as data conversion, data processing etc. and cannot be avoided in practice.

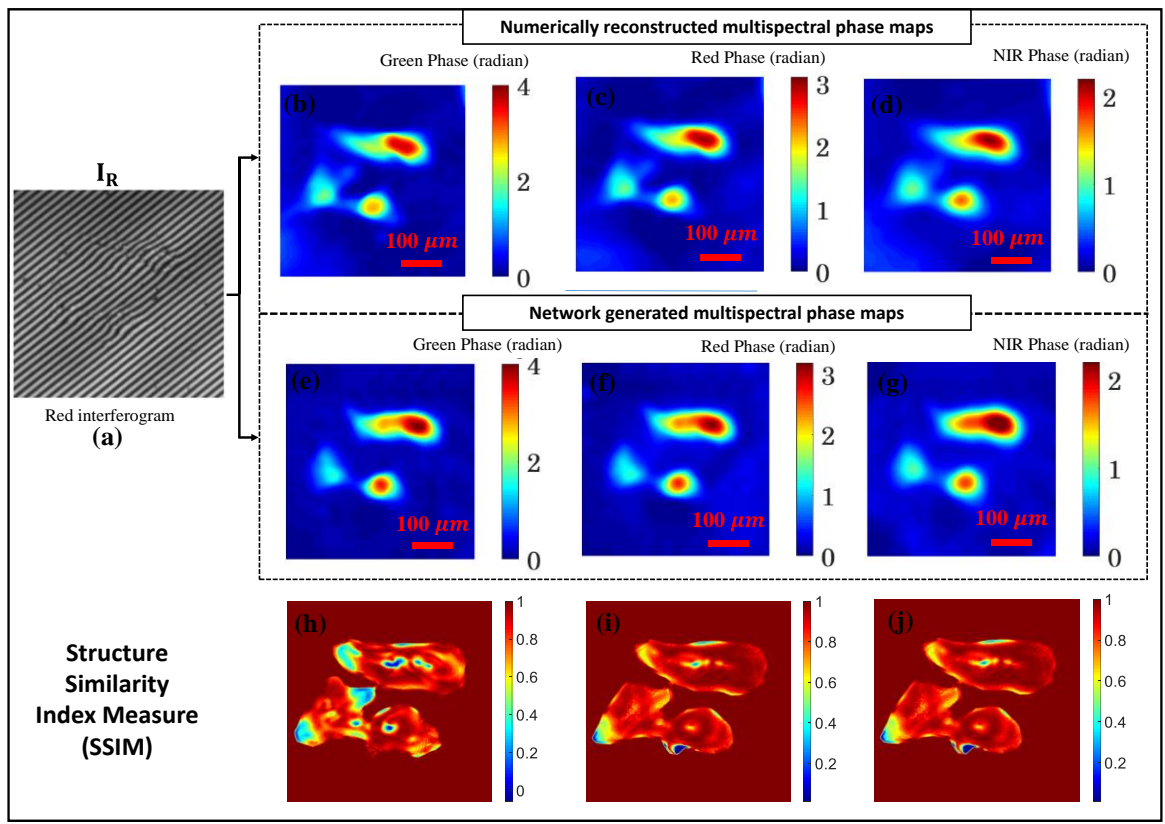

**FIGURE 5** Illustrating comparison between numerically (FT+ TIE algorithm based) reconstructed and network generated multispectral phase map data from a single input interferogram on MG63 osteosarcoma cells. (a) is the experimentally recorded input red wavelength interferogram. (b), (c) and (d) are the numerically reconstructed phase maps for experimentally recorded green, red, and NIR wavelength interferograms, respectively. (e), (f) and (g) are the network generated multispectral phase

maps corresponding to green, red, and NIR wavelengths, respectively. (h), (i) and (j) are the structure similarity index measure (SSIM) between numerically reconstructed and trained network generated multispectral phase maps. (h) represents SSIM for green wavelength phase maps between (b) and (e) which is found 0.94, (i) represents SSIM for red wavelength phase maps between (c) and (f) which is found 0.96 and (j) represents the SSIM for NIR wavelength phase maps between (d) and (g) which is found 0.96.

Figure 5 shows the comparison of numerically (FT+ TIE algorithm based) reconstructed and network generated multispectral phase map data from a single input interferogram on MG63 cell dataset. The N-IR wavelength $\lambda$ =808 nm is chosen since it has low absorption of biological molecules in this region[29]. Furthermore, NIR imaging is superior for thicker biological samples because it allows for better penetration at low illumination intensities, reducing the risk of photobleaching, achieving improved contrast and resolution, and reducing the possibility of cross-talk between channels. The input interferogram corresponding to $\lambda$=632.8 is shown in Fig. 5 (a). Figure 5 (b), (c) and (d) are numerically reconstructed phase maps corresponding to $\lambda$=532.8 nm, $\lambda$=632.8 nm and $\lambda$=808 nm, respectively for experimentally recorded interferograms. Figure 5 (e), (f) and (g) are the network generated phase maps corresponds to $\lambda$=532.8nm, $\lambda$=632.8nm and $\lambda$=808 nm, respectively. The comparison shows the close matching between network generated and experimentally reconstructed phase maps. Similar as in case of optical waveguide dataset, wavelength dependent phase map variation predicted by network. Further, the SSIM shown between the numerically reconstructed and network generated results validates the similarity in cell structures for MG63 cell dataset. The SSIM corresponds to $\lambda$=532.8nm between numerically reconstructed phase maps Fig. 5 (b) and network generated phase maps Fig. 5 (e) is 0.94, shown in Fig. 5 (h). Similarly, the SSIM corresponds to $\lambda$=632.8nm and $\lambda$=808nm between numerically reconstructed phase maps and network generated phase maps are 0.96 and 0.96, shown in Fig. 5(i) and (j), respectively. The SSIM in Fig. 5 (h), (i) and (j), signifies that the similarity in structure between the network generated and numerically reconstructed phase maps corresponds to $\lambda$=532.8nm, $\lambda$=632.8nm and $\lambda$=808nm, respectively. The phase maps comparison results and the SSIM values between numerically reconstructed and network generated phase maps signifies that the proposed MS-QPI+DNN framework is trained for MS-QPI with higher similarity in structure for multi spectral data generation.

## Conclusion:

In the present manuscript, we have demonstrated a multi-spectral quantitative phase imaging (MS-QPI) technique using deep neural network (DNN). Single input interferogram (632.8 nm (red)) is used to generate three multi spectral phase maps corresponding to $\lambda$=532.8 nm (green), $\lambda$=632.8 nm (red) and $\lambda$=808 nm (NIR)) through the use of DNN trained specifically for this purpose and for samples of a specific kind. The performance of the network is tested on the optical waveguide and MG63 osteosarcoma cells. The comparison of the results in Fig. 4 and 5 shows that the network generated wavelength dependent phase map variation is in good agreement with experimentally reconstructed phase maps. The current framework can be used as a label free mode for mapping refractive index dependence of the biological cells and tissues with less-time consumption, without multi-wavelength light sources and bulky optical setup. An additional advantage of the proposed work is to use NIR wavelength light along with visible light sources, provide to NIR phase imaging without employing costly IR cameras. Further, conventional color crosstalk problem of MS-QPI can be resolved by using MS-QPI+DNN framework. Our proposed method shows single-shot MS-QPI which overcomes the requirement of multispectral interferograms for multi spectral phase generation. Furthermore, no mechanical moving, multi data acquisition and processing is required so time consumption is reduced in MS-QPI. The present methods are useful for the QPI of biological cells and tissues such as human RBCs, and cancer tissues where label-free imaging is important in disease diagnosis and identification of cancer margin.

## Acknowledgement:

S.B. would like to acknowledge CSIR, India for financial support of this research work through the research fellowship. K.A. and A.B. acknowledges the European Research Council Starting grant (id 804233) and INTPART (id 309802). The author would like to acknowledge Dibakar Borah, "Department of Physics, IIT Delhi, India" for helping in the manuscript preparation.